# Gamifying Cyber Governance: A Virtual Escape Room to Transform Cybersecurity Policy Education

Khondokar Fida Hasan[¥*]; William Hughes[¥]; Adrita Rahman[γ]

[¥]University of New South Wales (UNSW), Canberra, ACT, 2601
[γ]Bangladesh University of Business and Technology (BUBT), Dhaka, Bangladesh
*Corresponding Author Email: fida.hasan@unsw.edu.au

## Abstract

Serious games are gaining popularity as effective teaching and learning tools, providing engaging, interactive, and practical experiences for students. Gamified learning experiences, such as virtual escape rooms, have emerged as powerful tools in bridging theory and practice, fostering deeper understanding and engagement among students. This paper presents the design, implementation, and evaluation of a virtual escape room tailored specifically for cybersecurity governance and policy education. Developed as a 3D immersive environment, the escape room simulates a virtual company scenario to facilitate risk-informed cyber policy development. It consists of three interactive zones, each offering distinct sets of scenario-based problems that target specific educational objectives. Through these zones, students analyze cybersecurity risks, match security frameworks, and draft appropriate policies, thereby developing critical thinking, decision-making skills, and practical cybersecurity competencies. The primary contribution of this work lies in its innovative integration of game-based learning and immersive technology to create robust, interactive learning materials that are also resilient to generative AI interventions, thereby maintaining academic integrity. Additionally, the escape room provides students with exposure to real-world cybersecurity scenarios in a virtual office environment that meets industry expectations. Results from a student survey indicated strong positive feedback, highlighting significant improvements in students' engagement, practical understanding, and enthusiasm toward cybersecurity governance and policy concepts, underscoring the effectiveness and potential of gamification in cybersecurity education.

## 1. Introduction

Serious games have emerged as powerful educational tools, significantly enhancing engagement, motivation, and learning outcomes across various disciplines [1]. Unlike traditional pedagogical methods, serious games offer immersive, interactive environments that encourage active participation, critical thinking, and problem-solving skills through realistic simulations. These games integrate educational objectives into gameplay, allowing learners to acquire and apply knowledge experientially [2]. As digital and technological advancements transform educational landscapes, serious games offer promising avenues for innovative teaching and learning approaches, particularly in complex and rapidly evolving fields such as cybersecurity [3, 4].

Within serious games, virtual escape rooms have gained considerable attention due to their effectiveness in fostering immersive learning experiences through interactive problem-solving, narrative-driven challenges, and time-constrained tasks. In cybersecurity education, escape rooms are especially beneficial, offering realistic simulations that help bridge theoretical knowledge with practical skills. This immersive format encourages active participation and critical thinking, fostering deeper conceptual understanding and practical competence. Moreover, virtual escape rooms demonstrate resilience against the threats posed by generative AI to traditional assessment methods like essays and reports, as they require dynamic problem-solving and real-time decision-making, which are challenging for AI to replicate convincingly [5, 6].

However, despite the recognized potential of virtual escape rooms, significant gaps remain in the specific application of these tools to cybersecurity governance and policy education. Existing teaching methodologies in this area frequently rely on abstract, lecture-based, or textual instructional approaches, leading to student disengagement and limited ability to apply theoretical knowledge practically [4] [7-11]. Additionally, few studies have empirically evaluated how effectively virtual escape rooms enhance student engagement and the practical application of cybersecurity, including cyber governance and policy principles [7]. Addressing this gap is critical to developing more effective educational methods in cybersecurity, where the ability to apply governance, policy, and risk management frameworks practically is essential for professional success.

This paper addresses this gap by introducing a virtual escape room specifically designed to teach cybersecurity governance and policy through immersive and engaging gameplay. The virtual escape room consists of sequentially unlocked challenges, requiring students to solve real-world governance problems, align policies with cybersecurity standards, and manage risk under realistic constraints. Employing a design-based research approach, this study evaluates the effectiveness of the escape room in improving student engagement, conceptual understanding, and practical skills in cybersecurity governance and policymaking. The primary contribution of this paper is twofold: it introduces a rigorously designed educational tool specifically tailored for cybersecurity governance and policy education, and it provides empirical evidence demonstrating the tool's effectiveness in overcoming the limitations of traditional pedagogical methods such as challenges in student's engagements and also today's challenge posed by generative AI-influenced educational environment.

The remainder of this paper is structured as follows: Section 2 presents a short review of the literature on game-based learning and serious games in cybersecurity education. Section 3 outlines our research methodology and game development process. Section 4 presents the implementation details of the escape room, including its architecture and learning integration. Section 5 discusses evaluation findings, examining student engagement and learning outcomes. Finally, Section 6 concludes with key insights and potential directions for future research.

## 2. Background Study

Traditional pedagogical approaches, often centered on passive lectures and abstract theoretical concepts, frequently fail to fully engage students or equip them with the practical skills necessary to navigate the dynamic cybersecurity landscape [1, 2]. This section systematically reviews existing literature on the use of serious games in education, with a specific focus on their application in cybersecurity, and explores the potential of educational games as innovative and impactful pedagogical strategies.

**Serious Games (i.e., Educational Escape Room) in Education: A Transformative Approach**

Serious games, defined as games designed for a primary purpose other than pure entertainment, have emerged as a promising pedagogical tool across diverse educational domains [3]. Rooted in constructivist learning theories and principles of experiential learning, serious games offer immersive and interactive environments that actively involve learners in the learning process. By providing engaging contexts for problem-solving, decision-making, and experimentation, serious games can foster deeper understanding and knowledge retention compared to traditional passive learning methods [4].

Research consistently demonstrates the positive impact of serious games on various educational outcomes. Meta-analyses and systematic reviews have highlighted their effectiveness in enhancing student motivation and engagement, improving knowledge acquisition and conceptual understanding, and developing critical thinking and problem-solving skills [3-5]. Furthermore, serious games can provide safe and risk-free environments for learners to practice complex skills and apply theoretical knowledge in simulated real-world scenarios. The inherent interactive and feedback-rich nature of games facilitates active learning and allows learners to learn from mistakes in a supportive and engaging manner [6].

The application of serious games within cybersecurity education is particularly pertinent given the complex, dynamic, and practical nature of the field. Cybersecurity concepts are often abstract and challenging to grasp through purely theoretical instruction, and the rapid evolution of cyber threats necessitates the development of adaptable and practically skilled cybersecurity professionals. Serious games offer a unique opportunity to bridge the gap between theory and practice by simulating realistic cybersecurity scenarios and challenges in an engaging and interactive format [7, 8].

While the research base on serious games, specifically in cybersecurity education, is still developing, emerging studies demonstrate their potential to enhance learning in this domain. For example, studies have explored using serious games to teach network security concepts, incident response procedures, and ethical hacking skills. These studies often report improvements in student engagement, knowledge retention, and perceived self-efficacy in cybersecurity tasks [9]. Furthermore, serious games can effectively simulate the adversarial nature of cybersecurity, allowing students to experience the dynamic interplay between attackers and defenders in a controlled setting [10].

However, it is important to acknowledge that the field of serious games in cybersecurity education is still relatively nascent. There is a need for more rigorous empirical research to comprehensively evaluate the effectiveness of different types of serious games, identify best practices for game design and implementation in cybersecurity contexts, and explore their long-term impact on skill development and career readiness. Furthermore, many existing serious game approaches in cybersecurity might not fully address the challenges of student disengagement and the need for truly immersive and practical learning experiences [3, 6, 7, 10, 11].

Educational escape rooms, a specific type of serious game, have gained significant traction in recent years as an innovative and engaging pedagogical approach across various disciplines. Drawing inspiration from recreational escape rooms, educational escape rooms are designed with specific learning objectives in mind, embedding educational content and challenges within the narrative and puzzles of the escape room experience. They typically involve teams of participants working collaboratively to solve a series of puzzles and challenges within a time limit to "escape" from a simulated environment or achieve a specific in-game objective [5, 11].

While the literature on educational escape rooms is rapidly growing, their application in technical fields like cybersecurity is still relatively underexplored. Emerging evidence suggests that escape rooms can be effective in teaching computer science concepts and engineering principles. The immersive and collaborative nature of escape rooms may be particularly well-suited for cybersecurity education by simulating realistic scenarios requiring students to apply their knowledge and skills in a dynamic and engaging context, mimicking the pressures and teamwork often encountered in cybersecurity incidents [7, 8].

**Theoretical Framework: Constructivist Learning Theory**

This research is underpinned by constructivist learning theory, which posits that learners actively construct knowledge through experience and interaction with their environment. Serious games, and educational escape rooms, in particular, align strongly with constructivist principles by providing learners with opportunities for [3, 12]:

- Active engagement: Learners are not passive recipients of information but actively involved in problem-solving and knowledge construction through game play.
- Experiential learning: Learners learn by doing, applying theoretical knowledge in simulated, practical scenarios within the escape room.
- Social interaction and collaboration: Escape rooms inherently promote teamwork and peer learning, allowing students to learn from each other and build knowledge collaboratively.
- Meaningful context: The narrative and game mechanics provide a meaningful and relevant context for learning cybersecurity concepts, increasing motivation and engagement.
- Feedback and reflection: Escape rooms provide immediate feedback through game progression and outcomes, allowing learners to reflect on their actions and refine their understanding.

By situating learning within an engaging and interactive escape room environment, this innovation seeks to leverage the principles of constructivism to foster deeper and more meaningful learning experiences for cybersecurity students.

## 3. Research Methodology and System Development

The system development of the cybersecurity governance escape room was guided by a structured methodology comprising three interconnected phases: Planning & Requirements, Process, and Output, as shown in Figure 1. This methodology extends beyond conventional serious game development frameworks [23] by addressing the distinctive challenges associated with cybersecurity education, particularly in governance and policy development [9]. Each phase is described in following.

### 3.1. Planning & Requirements Phase

The Planning & Requirements phase laid the foundation for our development process. We defined clear learning objectives focused on policy formulation, risk assessment, and governance principles that align with both academic standards and industry needs. Our audience analysis revealed varying levels of cybersecurity expertise and learning preferences, which informed our design of difficulty progression and support mechanisms. We ensured curriculum alignment by creating clear connections to existing coursework, complementing traditional teaching methods while establishing pathways for skill assessment. The game incorporates current cybersecurity governance frameworks to create authentic scenarios that bridge theory and practice, preparing students for real-world professional settings. For implementation, we selected virtual environment specifications that maximize accessibility while working within technological constraints. After assessing available resources, we adopted appropriate technical solutions for deployment. The game supports both individual and collaborative participation, with session durations designed to maintain engagement while achieving learning outcomes.

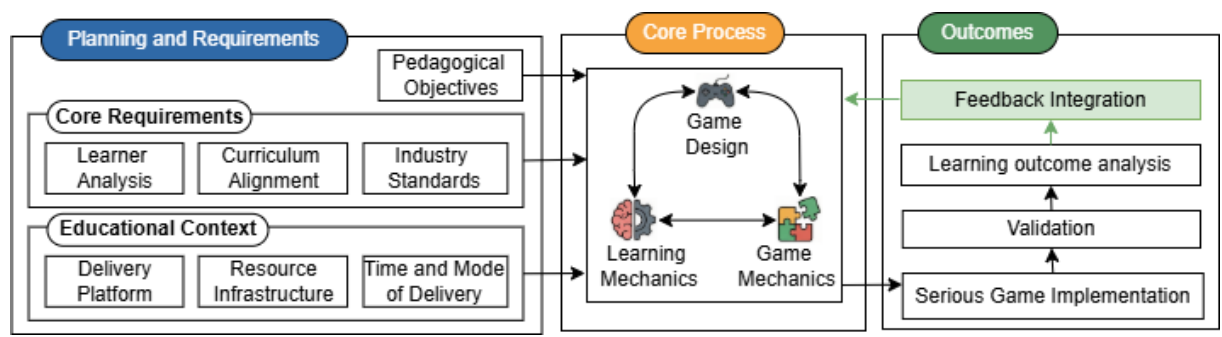

Figure 1 Methodology of Game Design, Development, Validation, and Feedback

### 3.2. Process Phase: Iterative Design Approach

The Process phase used an iterative approach where we developed and refined learning mechanics, game design, and game mechanics together. We created initial prototypes, tested them, gathered feedback, and made improvements in repeated cycles. This helped us effectively combine learning objectives with gameplay.

Our learning mechanics-built knowledge progressively, with students facing increasingly complex challenges that required deeper understanding of cybersecurity governance. The game design featured a narrative where each room represented a different aspect of security governance. These rooms connected both physically and conceptually, building on previous knowledge while introducing new concepts.

We designed game mechanics to support learning objectives while keeping students engaged. The escape room format naturally facilitated problem-solving, requiring students to identify security risks, analyze governance challenges, and develop appropriate policies. Through iteration, we refined how challenges were presented and how feedback was provided to maintain engagement and support learning.

### 3.3. Output Phase: Implementation and Validation

The Output phase marked our shift from development to implementation and evaluation. We integrated all refined components into a complete educational game with well-designed challenges, an intuitive interface, and effective assessment tools.

We validated the game through multiple methods. First, cybersecurity professionals reviewed the content for technical accuracy and relevance. Then, we conducted comprehensive testing to verify both technical functionality and educational effectiveness. Our assessment went beyond traditional knowledge tests to measure skill development and practical application.

## 4. Development of UNSW Cyber Escape Room

The UNSW Cyber Escape Room represents a game-based learning environment designed to enhance cybersecurity education in governance, policy, and risk management. The module targets students enrolled in cybersecurity courses, immersing them in a virtual office environment where they solve a series of progressive challenges to advance through the game.

### 4.1. Game Architecture Overview

The Escape Room module is a game-based learning experience designed to enhance cybersecurity education in the areas of governance, policy, and risk management. The module is targeted at undergraduate students enrolled in cybersecurity courses. It immerses students in a virtual office environment where they must solve a series of cybersecurity-related challenges to "escape" each room.

The game environment consists of several individual rooms, each representing a different aspect of cybersecurity governance, policy, or risk management. Students must work collaboratively to solve puzzles, decipher clues, and apply their knowledge of cybersecurity Governance and Policy concepts to progress through the game.

The architecture of the cybersecurity governance escape room was systematically designed to integrate pedagogical principles with immersive game mechanics. As illustrated in Figure 2, the architecture comprises two primary layers: the Scene Flow and the Game Engine.

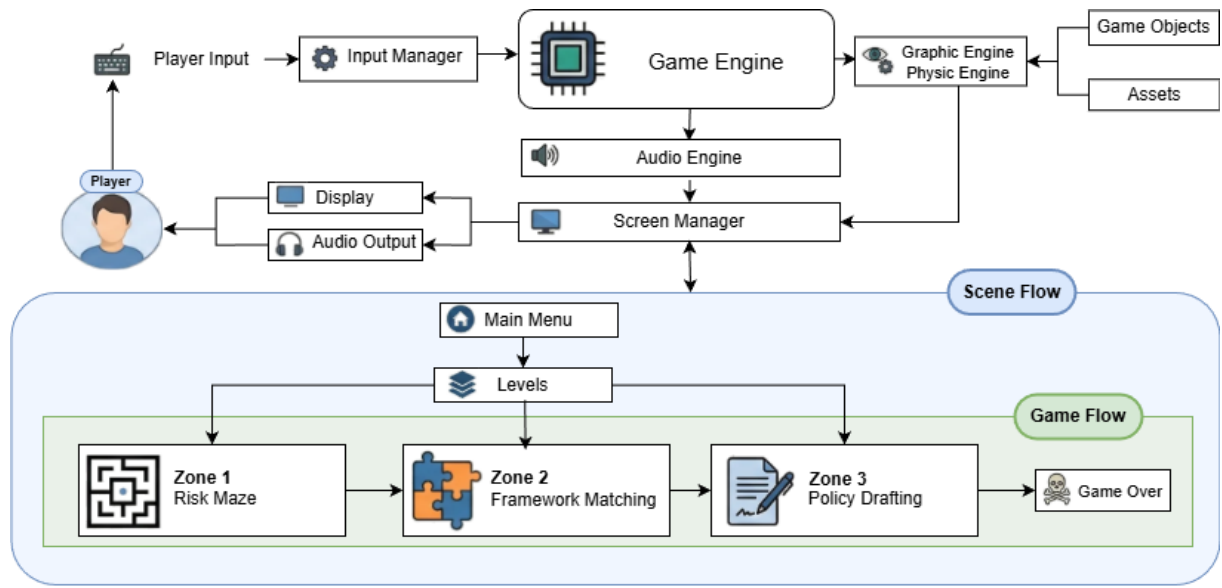

Figure 2 Architecture of the Developed Game

The Scene Flow governs the game's narrative structure and progression. It begins with a Main Menu, allowing players to navigate different levels. The Game Engine underpins the functional aspects of the game, handling both graphical and computational components. The engine includes the following subsystems:

- Graphic Engine: Responsible for rendering the virtual environment and visual elements.
- Audio Engine: Manages sound effects and auditory feedback, enhancing engagement.
- Scene Management: Governs different game environments, including player interactions, lighting, and UI components.
- Player Controller: Facilitates user input and interaction, allowing players to navigate the escape room effectively.

The Player Interaction Loop ensures real-time engagement, integrating user inputs with graphical and auditory feedback mechanisms. Players interact with the system through inputs, which are processed by the game engine, triggering responsive audiovisual outputs that drive gameplay progression. This architecture fosters an engaging yet structured learning experience, seamlessly integrating educational objectives with dynamic gameplay elements.

### 4.2. Problem Formulation and Game Mechanics

The Scene Flow governs the game's narrative structure and progression. It begins with a Main Menu (i.e. Landing Page), allowing players to navigate to know instructions on gameplay. The game consists of three sequential zones:

**Zone 1 - Risk Maze:** The Risk Maze zone focuses on developing students' ability to identify and prioritize cybersecurity risks in organizational contexts.

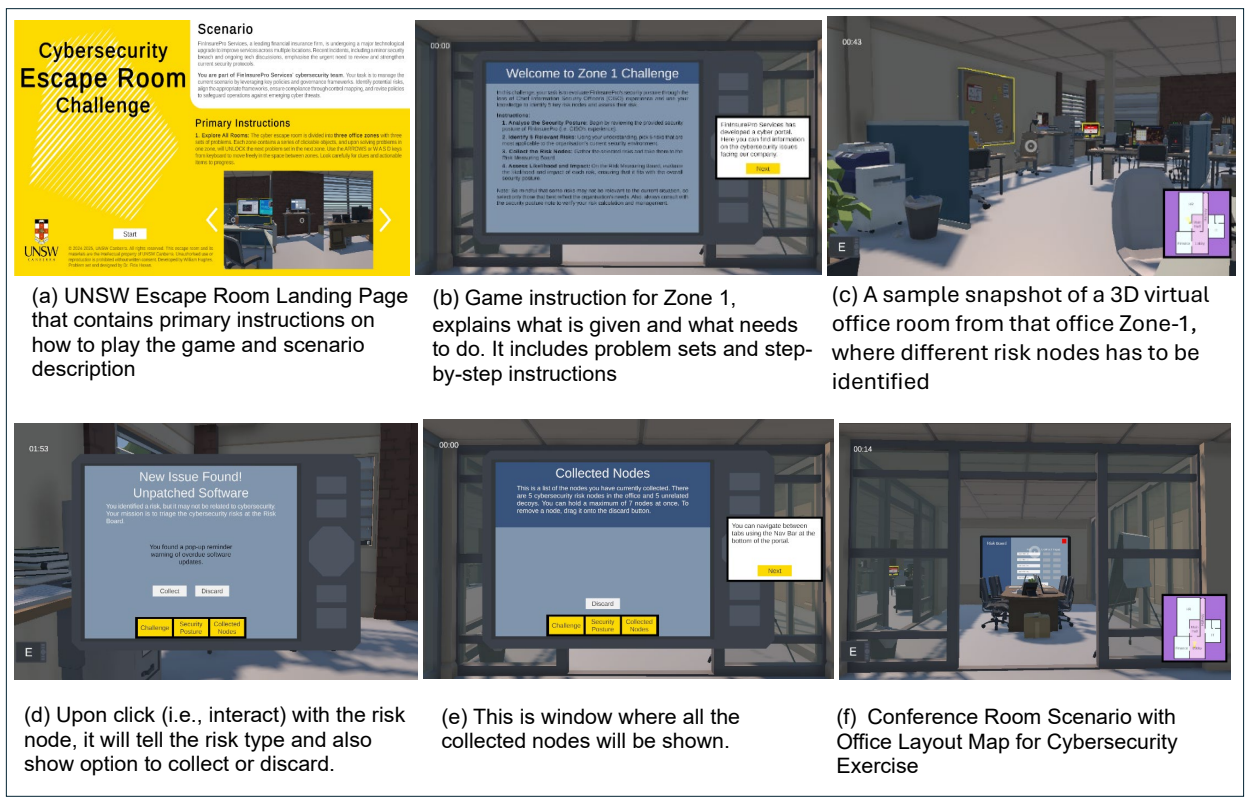

(a) UNSW Escape Room Landing Page that contains primary instructions on how to play the game and scenario description

(b) Game instruction for Zone 1, explains what is given and what needs to do. It includes problem sets and step-by-step instructions

(c) A sample snapshot of a 3D virtual office room from that office Zone-1, where different risk nodes has to be identified

(d) Upon click (i.e., interact) with the risk node, it will tell the risk type and also show option to collect or discard.

(e) This is window where all the collected nodes will be shown.

(f) Conference Room Scenario with Office Layout Map for Cybersecurity Exercise

*Figure 3 Snap of various stages of the developed UNSW Escape Room: Zone-1 Game Scenario and Game Mechanics*

The maze-like environment creates a metaphorical representation of navigating complex risk landscapes. As students move through the room, they encounter various risk scenarios that require risk identification in ambiguous situations, prioritization of competing security concerns, application of risk assessment methodologies, and decision-making on organizational context. The puzzles in this zone require students to demonstrate understanding of key risk management frameworks, translating theoretical knowledge into practical application.

**Zone 2 - Framework Matching**: The Framework Matching zone reinforces students' understanding of cybersecurity governance frameworks and compliance standards. This zone presents students with a series of security controls and requires them to match controls to appropriate frameworks (e.g., NIST CSF, ISO 27001, GDPR), identify overlapping requirements across different standards, recognize framework-specific terminology and requirements, and understand the contextual application of controls in different industries. This zone employs a combination of puzzle mechanics, including sorting, classification, and pattern recognition, to create an engaging learning experience that reinforces the relationships between different governance frameworks.

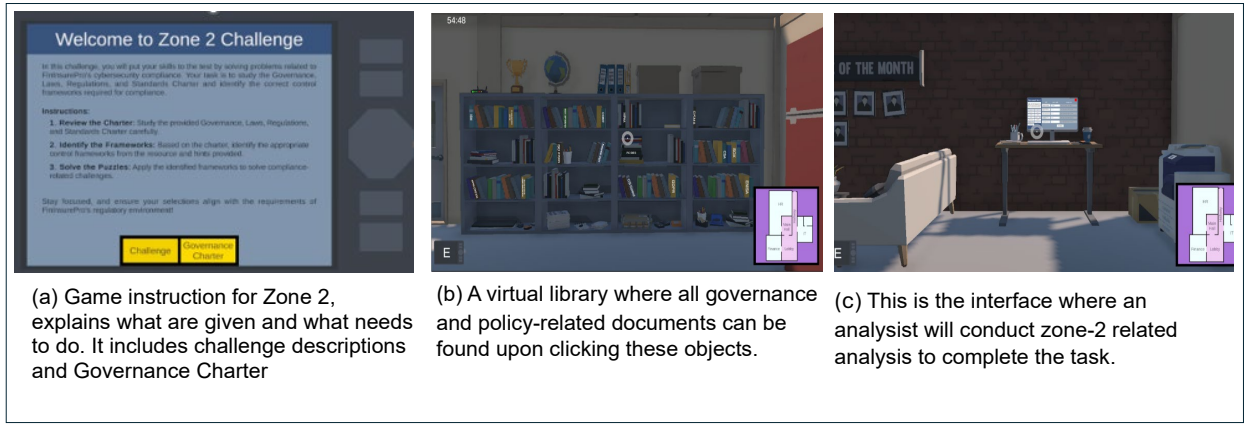

(a) Game instruction for Zone 2, explains what are given and what needs to do. It includes challenge descriptions and Governance Charter

(b) A virtual library where all governance and policy-related documents can be found upon clicking these objects.

(c) This is the interface where an analyst will conduct zone-2 related analysis to complete the task.

Figure 4  Snap of various stages of the developed game from Zone 2

**Zone 3 - Policy Revisioning and Drafting:** The Policy Drafting zone challenges students to synthesize their knowledge of risks and frameworks into actionable policy documents. This culminating zone requires higher-order thinking skills as students must revise the existing policy and then draft new policy statements that address identified risks, ensure policy alignment with relevant governance frameworks, consider stakeholder impacts and implementation constraints, and balance security requirements with organizational needs. The zone employs a structured policy template with interactive elements that guide students through the policy creation process while allowing creative problem-solving within governance constraints.

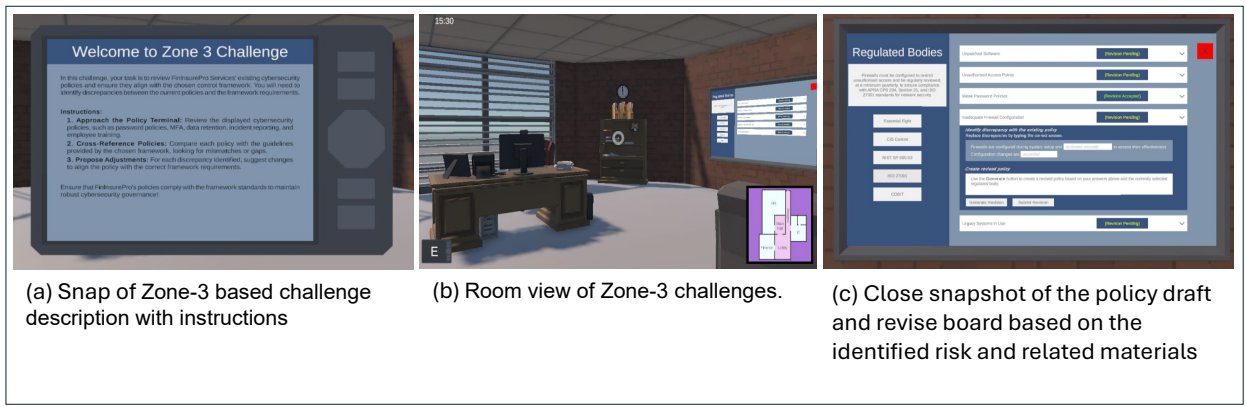

(a) Snap of Zone-3 based challenge description with instructions

(b) Room view of Zone-3 challenges.

(c) Close snapshot of the policy draft and revise board based on the identified risk and related materials

Figure 5  Snap of various stages of the developed game from Zone 3

The zone employs a structured policy template with interactive elements that guide students through the policy creation process while allowing creative problem-solving within governance constraints. Technically, this is implemented through a component-based system where students select and arrange policy elements from a pre-validated library of options, with the game engine evaluating policy completeness based on the inclusion of essential components such as scope, roles, and responsibilities, compliance requirements, and enforcement mechanisms. The system uses rule-based validation to ensure policies address the specific risks identified in Zone 1 and incorporate appropriate frameworks from Zone 2, creating a cohesive learning progression across all three zones. Real-time feedback is provided through visual indicators that highlight gaps or inconsistencies in the drafted policies, prompting students to refine their work until it meets the established criteria for effectiveness.

## 5. Evaluation and Results

To assess the effectiveness of the developed 3D serious game virtual escape room designed for teaching cybersecurity governance and policy, we conducted a comprehensive student survey. The evaluation aimed to measure student engagement, motivation, perceived difficulty, and overall learning effectiveness.

Student engagement was evaluated through structured survey questions. When asked, "How engaging did you find the Escape Room activity?", approximately 71% of students provided the highest rating (5), and an additional 21.4% rated their engagement level as 4 as shown in Figure 1. This suggests significant positive engagement with the escape room.

Furthermore, when participants were asked, "Did the interactive nature of the Escape Room increase your interest in learning cybersecurity concepts compared to traditional methods?", all responses were positive. A significant majority, 57.1%, strongly affirmed with "Oh Definitely," while 42.9% agreed with "Yes," as illustrated in Figure 2. This feedback highlights the effectiveness of interactive, game-based learning approaches in enhancing student motivation for learning complex topics such as cybersecurity governance.

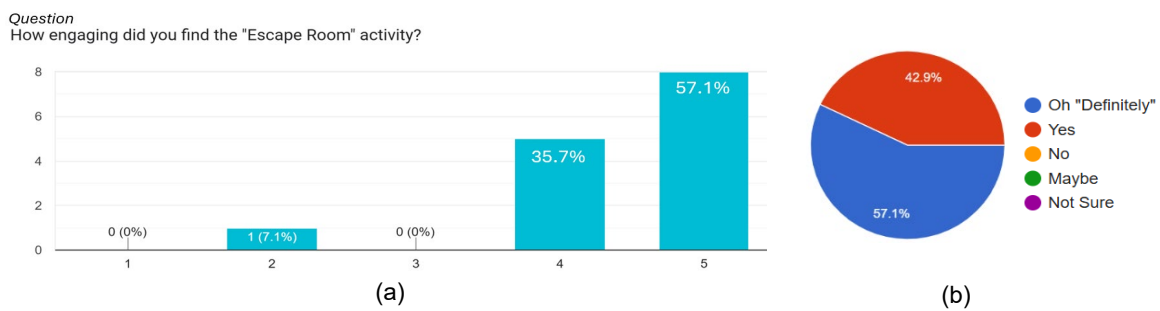

*Figure 6 Student's Survey on Engagement*

Additionally, overall satisfaction with the escape room experience was measured through the question, "On a scale of 1 to 5, how satisfied are you with the overall Escape Room learning experience?" Responses indicated notably high satisfaction, with 71% of students rating their satisfaction at the highest level (5), and 21.4% rating it as 4. Only a small minority (7.1%) expressed dissatisfaction with a rating of 1, as depicted in Figure 2.

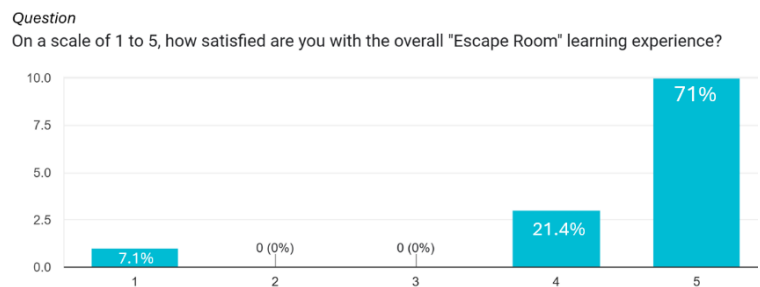

*Figure 7 Student's Survey on Overall Satisfaction*

Another important aspect of our evaluation focused on the difficulty calibration of the escape room challenges. Students were asked to assess the difficulty level of the activities, providing insights into whether the challenges were appropriately calibrated for their knowledge level and learning progression. The feedback regarding difficulty level suggests that the escape room achieved an appropriate balance between challenge and accessibility. Most students rated the difficulty in the mid-range of the scale, indicating that the activities were neither too simple to be engaging nor too complex to be discouraging. This balanced difficulty level is particularly important in educational games, where excessive difficulty can lead to frustration and disengagement, while insufficient challenge can result in limited learning outcomes. The progressive difficulty structure across the three zones (Risk Maze,

Framework Matching, and Policy Drafting) appears to have been effective in scaffolding students' learning experiences. This structured progression allowed students to build confidence with foundational concepts before engaging with more complex governance challenges, contributing to the overall positive engagement metrics [24].

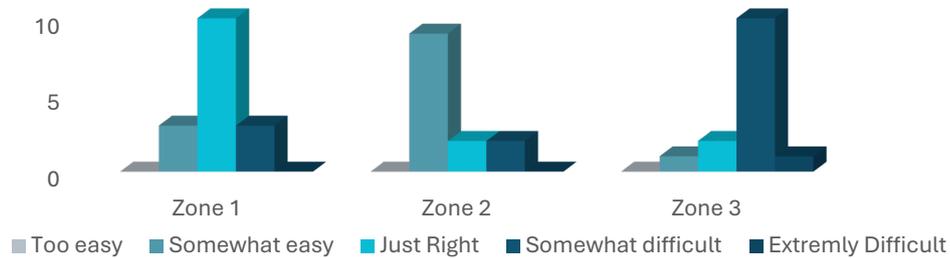

Figure 8 Student's Survey on Difficulty Calibration

Overall, the survey data indicates that the escape room format successfully addressed our core research question—how to enhance student engagement with governance and policy concepts. The strong positive response regarding increased interest in learning cybersecurity concepts demonstrates that the gamified approach effectively overcame engagement barriers associated with these topics.

While our evaluation is limited by the sample size and scope of feedback collected, the consistently positive responses suggest that the escape room format offers a promising approach to cybersecurity governance education. The interactive nature of the experience transforms abstract concepts into engaging learning activities that resonate with students while providing assessment opportunities resistant to generative AI circumvention through their dynamic, contextual nature.

## 6. Conclusion and Future Work

This paper introduces the UNSW Cyber Escape Room, an innovative virtual experience tailored for cybersecurity governance and policy education. Built as a 3D immersive environment, it realistically simulates a virtual company to guide students through risk-informed policy development. The escape room comprises three progressive zones: Risk Maze, Framework Matching, and Policy Drafting. Each zone presents interactive, scenario-based challenges designed to effectively connect theoretical cybersecurity concepts with real-world practical applications. Our evaluation shows notable improvements in student engagement, practical comprehension, and enthusiasm toward cybersecurity governance topics. The thoughtfully calibrated difficulty level ensures an engaging experience without overwhelming participants, promoting active involvement and critical thinking. Additionally, this game-based approach maintains academic integrity by offering assessments robust against generative AI interference. The methodological approach employed, integrating clear learning objectives, interactive gameplay, and iterative refinements, provides a flexible framework adaptable to diverse educational settings. Future research should explore detailed assessments of knowledge gains, skill progression, and long-term retention to further demonstrate the value and effectiveness of gamification in cybersecurity education.